
\magnification=1200



\def\eg{{\it e.g.,\ }}
\def\ie{{\it i.e.,\ }}



\def\cO{{\cal O}}
\def\cS{{\cal S}}


\def\a{\alpha}
\def\be{{\beta}}
\def\ga{{\gamma}}
\def\de{{\delta}}
\def\eps{{\epsilon}}
\def\veps{{\varepsilon}}
\def\ze{{\zeta}}

\def\si{{\sigma}}

\def\om{{\omega}}
\def\G{{\Gamma}}
\def\D{{\Delta}}

\def\Si{{\Sigma}}

\def\Om{{\Omega}}


\def\ref#1{Ref.~#1}			
\def\[#1]{[\cite{#1}]}
\def\cite#1{{#1}}

\def\slL{\raise.15ex\hbox{$/$}\kern-.53em\hbox{$L$}}
\def\slP{\raise.15ex\hbox{$/$}\kern-.53em\hbox{$P$}}
\def\slR{\raise.15ex\hbox{$/$}\kern-.53em\hbox{$R$}}
\def\slQ{\raise.15ex\hbox{$/$}\kern-.53em\hbox{$Q$}}
\def\slcD{\raise.15ex\hbox{$/$}\kern-.53em\hbox{${\cal D}$}}


\font\bx=cmr8
\font\section=cmbx10 scaled\magstep 2
\null
\pageno=0
\vskip -20pt
\line {\hfill Revised Version}
\vskip 2.5truecm
\centerline {\section Retarded/Advanced Correlation Functions }
\vskip0.4truecm
\centerline{\section and soft photon production }
\vskip0.4truecm
\centerline{\section in the Hard Loop Approximation }
\vskip 1.5truecm
\centerline {P. Aurenche{\footnote{\bx $^\dagger$}{\baselineskip=9pt\bx
Institute  for Theoritical Physics, University of California, Santa
Barbara, CA 93106-4030, USA}}$^{,\ddagger}$,
T. Becherrawy$^{\ddagger,}${\footnote{\bx $^\star$}{\baselineskip=9pt\bx
Laboratoire de Physique Quantique, Facult\'e des Sciences, BP 239, F-54506
Vandoeuvre-les-Nancy}}}
\vskip 0.4truecm
\centerline {and}
\vskip 0.4truecm
\centerline {{E.Petitgirard}{\footnote{\bx $^\ddagger$}{\baselineskip=9pt\bx
Laboratoire de Physique Th\'eorique ENSLAPP (URA 14-36 du CNRS, associ\'e
\`a l'E.N.S. de Lyon, et au L.A.P.P. (IN2P3-CNRS) d'Annecy-le-Vieux) Chemin
de Bellevue, BP 110, F-74941 Annecy-le-Vieux Cedex, France}}}
\vskip 0.4truecm
\vskip 0.4truecm
\centerline {\bf Abstract}
\vskip 1truecm
We apply the retarded/advanced formalism of real time field theory to the QED
or QED
like case. We obtain a general expression for the imaginary part of the
two-point correlation function in terms of discontinuities.The hard loop
expansion is derived. The formalism is used to extract the divergent part of
the soft fermion loop contribution to the real soft photon production.
\vskip 1.5truecm
\line {\hfill ENSLAPP-A-452/93}
\line {\hfill NSF-ITP-93-155}
\line {\hfill December 1993}
\vfill\eject

\noindent
{\bf I. {Introduction}}

\vskip .50cm

In the study of field theory at high temperature [\cite{1}], the development of
resummation techniques [\cite{2-4}] has considerably extended the domain of
validity of the perturbative approach. From the practical point of view,
several possible signals for the formation of the quark-gluon plasma have now
been studied in this framework. In particular the production of soft virtual
photons [\cite{5,6}] as well as hard real photons [\cite{7,8}] have been
discussed in great detail. Besides providing a consistent perturbative
expansion, the resummed series gives a finite result for quantities which would
otherwise not be well defined in a more naive approach.

The resummation technique of ``hard loop" expansion has been formulated in the
imaginary-time formalism (ITF) [\cite{1,9,10}] of thermal field theories. On
the other hand, there has been many developments in the real-time formalism
(RTF) [\cite{1,10-12}] or in the thermo-field dynamics (TFD) approach
[\cite{13}]. The latter seems more appropriate for the study of dynamical
phenomena since time is kept as an independent variable throughout the
calculation. Equivalently, in momentum space, no analytic continuation has to
be performed to obtain physical quantities.
Several studies have been devoted to the comparison of the ITF and the RTF
[\cite{14-20}]. In particular, a simple way to relate the two formalisms
consists in constructing in the RTF the retarded/advanced (R/A) Green's
functions which, at least for the two- and three-point functions, have a close
connection to the ITF Green's functions. In this paper, we pursue the study of
the R/A functions and show, in QED, how to formulate the hard-loop expansion:
several formulae are derived which are useful in the calculation of physical
quantities.

In the R/A formulation of the RTF, the propagators are still $2\times2$
matrices but they are defined to be diagonal and constructed, at least at the
lowest order of perturbation theory, from the retarded and advanced
propagators of the $T=0$ theory [\cite{17,18}].
As for the vertices, they become temperature dependent.
The R/A (amputated) Green's functions are in fact related to the causal
combinations of RTF Green's functions first introduced by Kobes [\cite{14,15}].
In TFD, which has very similar Feynman rules to the RTF, such a
diagonalization can also be performed and it is related to a thermal Bogoliubov
transformation, thereby allowing the interpretation of the diagonal
propagators as those of statistical quasi-particles [\cite{21}]. This
formulation seems therefore quite naturally appropriate to the study of thermal
field theories.

In the following, we first complete the study started in [\cite{17}] by
constructing explicitly the diagonalization for fermion propagators and
defining the Feynman rules in the R/A approach.
We then discuss, in QED, the one loop expression for the 2-, 3- and 4-point
functions in the context of the hard loop approximation and we easily recover
the ITF results continued to real energies. We then set up a general formula,
at the multi-loop level, for the 2-point functions and discuss how it is
related to the ITF results. As an application, we consider the production of
soft real photons and show that, despite cancellations due to thermal gauge
invariance between various terms of the hard loop expansion, there survives a
divergence. One of the advantages of the R/A approach over the ITF one is that
it does not require any analytical continuation. Furthermore the spin
structure of the diagrams is
\vfill
\eject
\noindent
the same as at $T=0$ and no continuation from Euclidian
to Minkowski space is needed.
Also, the possibility of using contour integration
may simplify the calculation.

\vskip .50cm
\noindent
{\bf II. The propagators and vertices in the R/A formalism.}

\vskip .50cm

In the real-time formalism, the fermion propagator, defined on a contour
characterized by $\sigma$ (fig.~1) can be written
as a product of $2\times2$ matrices [\cite{17}].
$$
S_F(P)=(\slP+M)\ U^F (P)\ \tilde D(P)\ V^F(P). \eqno(1)
$$
Denoting the retarded and advanced bosonic propagators at 0 temperature
$$
\eqalign{
\Delta_R (P) & = {i\over P^2-M^2+i\veps p_0} = {i\over 2 \Om}
 \left( {1\over p_0-\Om+i\veps} - {1\over p_0+\Om+i\veps}\right) \cr
\Delta_A (P) & = {i\over P^2-M^2-i\veps p_0} = {i\over 2 \Om}
\left({1\over p_0-\Om-i\veps}-{1\over p_0+\Om-i\veps}\right)\cr}, \eqno(2)
$$
$(\Om=({\vec p}^2 + M^2)^{1/2})$ we introduce the diagonal matrix
$$
\tilde D (P) = (\tilde D_{\a\be})= \pmatrix{\Delta_R(P) & 0 \cr
					0 & \Delta_A(P) \cr} \eqno(3)
$$
The other matrices will be specified shortly after we define the photon
propagator. In the Feynman gauge, the latter takes the form
$$
G^{\mu \nu}(P)=-g^{\mu \nu}\ U^B (P)\ \tilde D (P)\ V^B (P) \eqno(4)
$$
where the same diagonal matrix as in eq.~(1) appears. The diagonalization
matrix, denoted generically $U^{[\eta]},\ V^{[\eta]}$, are defined in a similar
fashion for bosons and fermions, and besides the contour $\si$, they depend on
arbitrary scalar functions $b(p),\ c(p)$
$$
\eqalignno{
U^{[\eta]}(P) & = (U^{[\eta]}_{i\a}) =-\eta \  n^{[\eta]} (-p_0)
	\pmatrix{b^{-1} & \eta c^{-1} e^{(\si-\be)p_0} \cr
		 b^{-1} e^{-\si p_0} & c^{-1} \cr} & (5) \cr
V^{[\eta]}(P) & = (V^{[\eta]}_{\a i}) = \pmatrix{b & \eta b e^{(\si-\be)p_0}\cr
					-c e^{-\si p_0} & -c\cr} & (6) \cr}
$$
where, not surprisingly, $\eta=1\ ([\eta]=B)$ for a boson and
$\eta=-1\ ([\eta]=F)$ for a fermion
and $n^{[\eta]}(p_0)$ is the usual Bose-Einstein or Fermi-Dirac
distribution. $\be$ is the inverse of the temperature. The chemical potential
has been set to 0, but it is easily introduced by the substitution $\be p_0\to
\be (p_0-\mu)$, $\si p_0$ remaining invariant.

The basic principle, in the R/A formalism, is to associate the diagonalization
matrices to the vertices in a natural way, keeping therefore the matrices
$\tilde D$ as diagonal propagators. This leads to different types of vertices
depending on the momentum flow. For instance, for all incoming momenta as in
fig.~2a $(P+Q+R=0)$ we introduce
$$
-i\ga_{\a\be\de} (P,Q,R)=-ig_{abd}\ V^F_{\a a}(P)\ V^B_{\be b}(Q)
\ V^F_{\de d}(R) \eqno(7)
$$
The latin indices refer to the 1 (particle) or 2 (ghost) fields of the usual
formulation of RTF so that $g_{111}=e,\ g_{222}=-e$ ($e$ is the electron
charge), all the other couplings being 0. The greek indices take the values R
or
A. In defining the vertex function $\ga_{\a\be\de}$ we leave out the Dirac or
Lorentz part of the vertex and keep only its scalar R/A structure.
{}From the definitions above it can be shown that
$$
\eqalign{
\ga_{\a\be\de}(P,Q,R) & =e\ (b(P))^{\de_{\a R}}\ (b(Q))^{\de_{\be R}}\
 (b(R))^{\de_{\de R}}\cr
& (-c(P))^{\de_{\a A}}\ (-c(Q))^{\de_{\be A}}\ (-c(R))^{\de_{\de A}}\
e^{\si L_0}\ (1-(-1)^{\de_{\a R}+\de_{\de R}} e^{-\be L_0}) \cr} \eqno(8)
$$
with $L_0=p_0\de_{\a R}+q_0\de_{\be R} + r_0 \de_{\de R}$.

It is immediately clear at this point that $\ga_{AAA}$ always vanishes while
$\ga_{RRR}=0$ because of momentum conservation. These results reflect the
causality requirement that three particles propagating forward in time (or
backward in time) cannot annihilate into (or be created from) the vacuum. It is
convenient, and perhaps more natural to introduce the vertex in figure~2b,
where the flow of momentum follows the fermion line. Momentum conservation now
reads $P+Q=R$. We define
$$
-i\ga_{\a\be;\de} (P,Q,R)=-ig_{abd}\ V^F_{\a a}(P)\ V^B_{\be b} (Q)
\ U^F_{d\de} (R) \eqno(9)
$$
An expression, similar to eq.~(8), but not so symmetrical can be derived
$$
\eqalign{
\ga_{\a\be;\de} (P,Q,R) & =- e\ (b(P))^{\de_{\a R}}\ (b(Q))^{\de_{\be R}}
\ (b(R))^{-\de_{\de R}} \cr
& \ \  (-c(P) e^{-\si p_0})^{\de_{\a A}}\ (-c(Q) e^{-\si q_0)})^{\de_{\be A}}
\ (-c(R) e^{-\si r_0})^{-\de_{\de A}} \cr
& \ \  n^F(r_0) e^{\be  r_0 \de_{\de R}}\ \left[ (-1)^{\de_{\de R}} +
(-1)^{\de_{\a R}}
e^{-\be(p_0\de_{\a R}+q_0\de_{\be R}-r_0\de_{\de A})}\right] \cr} \eqno(10)
$$
Admittedly, this expression is not very illuminating, but it immediately
allows, by comparison with eq.~(8) to derive the relations
$$
\eqalign{
\ga_{\a\be; R} (P,Q;R) & =- {n^F(-r_0)e^{-\si r_0}\over b(R)c(-R)}\ \ga_{\a\be
A}
(P,Q,-R) \cr
\ga_{\a\be;A}(P,Q;R) & =- {n^F(r_0)e^{\si r_0}\over b(-R)c(R)}\ \ga_{\a\be R}
(P,Q,-R) \cr} \eqno(11)
$$
The obvious choice
$$
b(-R)\ c(R)=-n^F(r_0)\ e^{\si r_0} \eqno(12)
$$
simplifies the ``crossing" relation when the fermion momentum $R$ is changed to
$-R$, since we can then simply write
$$
\ga_{\a\be;\de} (P,Q;R)=\ga_{\a\be\bar\de} (P,Q,-R) \eqno(13)
$$
where we have introduced $\bar\de=A,R$ the conjugate index of $\de=R,A$.
Similarly, if the ``crossing" property of the photon line is studied, we find
expression such as eq.~(11) with the difference that $n^B(-q_0)$ appears rather
than $-n^F(-q_0)$. The choice, in the bosonic diagonalization matrices
$$
b(-R)\ c(R)=n^B (q_0)\ e^{\si q_0} \eqno(14)
$$
allows to keep the same crossing property eq.~(13) for both fermion or boson
lines. Further choices can be made to simplify the expression of the thermal
vertices. For example, taking $b\equiv1$, we immediately obtain
$$
\eqalign{
\ga_{RRA} (P,Q,R) & = \ga_{ARR} (P,Q,R) =\ga_{RAR} (P,Q,R) =e \cr
\ga_{RAA}(P,Q,R) &=-e\ {n^B(q_0)\ n^F(r_0)\over n^F(q_0+r_0)} =-
e\ (1+n^B(q_0)-n^F(r_0)) \cr
\ga_{ARA} (P,Q,R) & =- e\ {n^F(q_0)\ n^F(r_0)\over n^B(q_0+r_0)} =-
e\ (1-n^F(q_0)-n^F(r_0)). \cr} \eqno(15)
$$
Together with
$$
\ga_{RRR} = \ga_{AAA} = 0 \eqno(16)
$$
and the temperature independent propagators
$$
\eqalign{
\tilde S_F (P) & = (\slP+M) \pmatrix{\D_R(P) & 0 \cr 0 & \D_A(P)\cr} \cr
\tilde G^{\mu \nu}(P) & =- g^{\mu \nu}
\pmatrix{\D_R(P) & 0 \cr 0 & \D_A(P)\cr} \cr}
\eqno(17)
$$
the diagrammatic rules for thermal QED in the R/A formalism are completely
specified. Let us note that all reference to the arbitrary contour $\si$ has
now disappeared from the Feynman rules. Unlike eqs.~(16), the results eqs.~(15)
and (17) are not universal. For instance, in [\cite{18}] the propagators are
chosen to be anti-diagonal so that reversal of momentum yields $\tilde D
(-P)= (\tilde D(P))^T$. Here we keep the diagonal form as it appears more
natural for the construction of loop diagrams. Furthermore, by a different
choice of the arbitrary functions $b$ and $c$, we could have imposed that the
vertices of type $\ga_{RAA}$ be equal to the coupling $e$. The vertices such as
$\ga_{RRA}$ would then have become linear in the statistical weights.

Any (amputated) $n$-point Green's function in the R/A formalism can be
constructed from the corresponding ones in the RTF by a generalization of
eqs.~(7) and (10). The R/A functions then appear as specific linear
combinations of the usual real-time functions. It is more practical, however,
if
one is interested in higher order perturbative calculations to construct the
R/A functions directly by application of the rules given above. Despite a
rather cumbersome notation, the Green's functions are nicely expressed in terms
of tree diagrams.

The diagonalization approach to RTF is more than an algebraic trick. The R/A
functions are ``more natural" than the Green's functions in the real-time
formalism. For example, it has been shown that the imaginary part of the
2-point function (there exists only two such functions related by complex
conjugation) is proportional to the opacity factor in the Boltzman equation for
a distribution near equilibrium [\cite{14,18}]. Furthermore, the $n$-point
functions with all but one indices set equal to $R$ are the causal Green's
functions introduced by Kobes [\cite{14,15}]. A different perspective comes
from thermo-field dynamics. The Feynman rules in TFD and RTF are very
similar (for systems in equilibrium which is the case considered here).
Recently
Henning and Umezawa [\cite{21}] studied the diagonalization of the 2-point
function in TFD. The diagonalization matrices (Bogoliubov transformations) have
two parameters: $\a$ which characterizes the thermal vacuum and $s$, with the
relations to our free parameters given by
$\a=1-\si/\be$ and $s=\ln b\sqrt{1+n(p_0)}=-\ln c\sqrt{1+n(p_0)}$. In TFD,
the Bogoliubov matrices parametrize the transformations which take the
physical (point-like) particles to the (thermal) quasi-particle
states and the diagonal (un-amputated) 2-point function $\tilde D(P)$ describes
the propagation of the quasiparticles in the medium. We note also that by
a differerent Bogoliubov transformation we could introduce the diagonal
matrix constructed with the usual Feynman propagator and its complex conjugate.

\vskip .30cm

We turn now to the study of loop diagrams.

\vskip .50cm
\noindent
{\bf III. R/A functions in the one-loop approximation}

\vskip .50cm

Following the method introduced for $\lambda \phi^3$ we
construct 2-, 3- and 4-point
functions perturbatively. We implicitly assume that we deal with QED or QCD. In
general, we do not specify the spin structure of the diagrams since it
factorizes from the R/A functions and it is the same as at 0 temperature. We
first consider a generic two-point function with external momentum $Q$ and
index $\be$ flowing
through the diagram. The internal momenta are $P$, with fermion of boson
statistics $\eta_P$, and $R$, with statistics $\eta_R$, such that $P+Q=R$
[fig.~3]. Applying the rules eqs.~(15)-(17) and using the crossing relation
eq.~(13) we arrive at
$$
\eqalign{
-i\Gamma_{\be\be} (Q) & =- e^2 \int {d^nP\over(2\pi)^n}
\ \slcD\  \ga_{\a\be\bar\de}(P,Q,-R)\ \Delta_\de (R)
\ \ga_{\bar\a\bar\be\de} (-P,-Q,R)\ \D_\a(P)\cr
&=- e^2\int {d^nP\over(2\pi)^n}\ \slcD\ \Bigg[ ({1\over2}+
\eta_P n^{[\eta_P]}(p_0))\ (\D_R(P)-\D_A(P)) \ \D_\be (R) \cr
&\ \ \ \  \ \ \  \ \ \ \ \ +({1\over2}+\eta_R n^{[\eta_R]}(r_0))
\ (\D_R(R)-\D_A(R))\ \D_{\bar\be}(P)\Bigg]. \cr} \eqno(18)
$$
The symbol $\slcD$
 denotes the Dirac and spin structure of the diagram. For example,
for the fermion self-energy we have, in Feynman gauge
$\slcD =-\ga_\nu\ (\slR+M)\ \ga^\nu$ if $R$ is the fermion internal line.
In arriving at the final form of eq.~(18) we have dropped in the integrand
terms of type $\D_R(P)\D_R(R)$ or $\D_A(P)\D_A(R)$ independent of the
statistical weight. These terms have poles in the $p_0$ complex energy plane
only on one side of the real axis. By closing the $p_0$ integration contour in
the other half-plane they are seen to give a vanishing contribution to the
2-point function. By the same token, the factor ${1\over2}$ can be replaced by
any constant because shifting the numerical value only generates irrelevant
$\D_R\D_R$ or $\D_A\D_A$ factors. We prefer keeping the ${1\over2}$ factor
since, using the relation
$$
\D_R(P)-\D_A(P) = 2\pi\ \veps(p_0)\ \de (P^2-M^2), \eqno(19)
$$
we find that it leads to an invariant form under the reversal of the sign of
momentum $P$:
$$
\left({1\over2}+\eta_P n^{[\eta_P]}(P)\right) \left(\D_R(P)-\D_A(P)\right)
=2\pi \left({1\over2} +\eta_P n^{[\eta_P]}(|p_0|)\right) \delta(P^2-M^2).
\eqno(20)
$$
When evaluating eq.~(18) we can either use the $\de$-function constraint or
close the $P_0$ contour in a conveniently chosen half-plane. For instance for
the first term in $\Gamma_{RR}(Q)$ encircling poles in the upper half-plane it
is seen that only the poles of $\D_A(P)$ contribute. In particular, the poles
on the imaginary axis associated to the statistical function never contribute
since they have a vanishing residue as $\D_R(P)-\D_A(P)=0$ on the imaginary
axis. For the second term, it is preferable to close the contour in the
lower-half plane to retain only the singularities of $\D_R(R)=\D_R(P+Q)$.

The two-point function obeys some general relations [\cite{17,18}]. We have
$$
\G_{\a\a}^*(Q) = \G_{\bar\a\bar\a}(Q) \eqno(21)
$$
provided, it is assumed that the complex conjugation does not operate on the
spinor structure $\slcD$ but only on the R/A part of the integrand. It is
based on the property
$$
\D^*_\a (P)=-\D_{\bar\a} (P). \eqno(22)
$$
We can also prove, for QED/QCD like theories and assuming massless particles,
that
$$
\G_{\a\a}(-Q)=\pm\G_{\bar\a\bar\a} (Q) \eqno(23)
$$
where the $(+)$ sign refers to a bosonic external line and the $(-)$ sign to a
fermionic one. This equation can be obtained by reversing the sign of all
momenta (external as well as internal) and using
$$
\eqalignno{
\D_\a(-P) & = \D_{\bar\a}(P) & (24)\cr
{1\over2} +\eta_P n^{[\eta_P]} (-p_0) & =- ({1\over2}+\eta_P n^{[\eta_P]}(p_0))
. & (25) \cr}
$$
Eqs.~(21) and (23) can be shown to hold at any loop order. Let us remark as a
consequence that, for massless fields, eq.~(23) holds true for the  full
propagators \ie for propagators after the geometrical series of self energy
corrections has been summed.

The other two point functions satisfy the obvious relation
$$
\Gamma_{\a\bar\a} (Q) =0 \eqno(26)
$$
which expresses the fact that a particle propagating forward (backward) in time
cannot turn into a particle propagating backward (forward) in time by
self-interactions. Technically, eq.~(26) is satisfied because its integrand is
a sum of terms having poles only one side of the real axis with no statistical
weights attached to them.

\vskip .50cm

Turning now to the 3-point function with the momenta as indicated in fig.~4
$(P+Q+R=0)$ we can derive by the same method as above
$$
\G_{\a\a\a}(P,Q,R) =0 \eqno(27)
$$
which is true to all order of perturbation theory. For the other functions we
find it convenient to factorize out the tree vertex expression of the 3-point
function and define (recall that $\ga_{\a\be\de}$  contains only the scalar
part of the vertex as defined in eq.~(15))
$$
V_{\a\be\de} (P,Q,R) = {\G_{\a\be\de} (P,Q,R)\over \ga_{\a\be\de}(P,Q,R)},
\eqno(28)
$$
not all indices being identical. It can be proven
$$
\eqalign{
V_{\a\be\de} & (P,Q,R) =- e^2 \int {d^nL_1\over(2\pi)^n} \ \slcD \cr
&\qquad \qquad \qquad \Bigg[({1\over2}+\eta_1n^{[\eta_1]}(l_{10}))
\ (\D_R(L_1)-\D_A(L_1)) \ \D_\a(L_2)\ \D_{\bar\de}(L_3) \cr
&\qquad \qquad \qquad + ({1\over2}+\eta_2n^{[\eta_2]}(l_{20}))
\ (\D_R(L_2)-\D_A(L_2)) \ \D_\be(L_3)\ \D_{\bar\a}(L_1)\cr
&\qquad \qquad \qquad + ({1\over2}+\eta_3n^{[\eta_3]}(l_{30}))
\ (\D_R(L_3)-\D_A(L_3)) \ \D_{\de}(L_1)\ \D_{\bar\be}(L_2)\Bigg]. \cr}
\eqno(29)
$$
The symbol $\slcD$ is now the appropriate one for the 3-point function.
Again, the
constant factor $1/2$ can be arbitrarily changed. We observe that eq.~(29)
appears as a sum of tree amplitudes since each $\D_R-\D_A$ combination puts the
corresponding line on shell with each cut line carrying a weight
$({1\over2}+\eta_i n^{[\eta_i]}(l_{i0}))\ \veps (l_{i0})\ \de(L^2_i-M^2)$ or
$\eta_i n^{[\eta_i]}(l_{i0})\ \veps (l_{i0})\ \de(L^2_i-M^2)$
if the factor $1/2$ is dropped in eq.~(29).
A graphical representation of this is given in fig.~5. As for the 2-point
function we derive [\cite{17,18}]
$$
\eqalignno{
V^*_{\a\be\de}(P,Q,R) & = V_{\bar\a\bar\be\bar\de}(P,Q,R) & (30)\cr
V_{\a\be\de}(-P,-Q,-R) & =\pm V_{\bar\a\bar\be\bar\de} (P,Q,R). & (31) \cr}
$$
Where the last equation holds true for massless fields in the QED/QCD case.
The $+$ sign is appropriate for a fermion-antifermion-gauge boson coupling,
while the $-$ sign is for the triple gauge boson coupling.

Turning to the 4-point functions they can also be expressed in terms of tree
diagrams. They are easily obtained from [\cite{17}] with the appropriate
modifications of the statistical factors, and taking into account the
appropriate Dirac structure.

\vskip .50cm
\noindent
{\bf IV. R/A functions in QED and the hard loop expansion.}

\vskip .50cm

In this section we write explicitly the relevant $n$-point functions in QED
and discuss Ward identities as well as the hard loop approximation.

After performing the loop energy integration in an $n$-point function, by use
of the $\de$-function or by a contour deformation, the temperature dependent
terms take the form
$$
\eqalign{
 \int {d^nL\over(2\pi)^n} \ n(l_0) & \ (\D_R(L)-\D_A(L))
\ F(l_0,\vec l;P,Q,\dots) \cr
& = {1\over2} \int{\om^{n-3}\over(2\pi)^{n-1}}\ n(\om) \ d \om \ \widehat{dl}
\ (F(\om,\vec l;P,Q,\dots) +F(-\om, \vec l;P,Q,\dots)) \cr} \eqno(32)
$$
where $\om=|\vec l|$ and $\widehat{dl}$ is the symbol for angular integration
in $n$-dimensions $(\hat l=\vec l/\om)$. We have assumed massless particles for
simplicity but the following argument also holds true in the massive case.
Keeping only relevant terms we have to evaluate in practical cases dimensional
integrals of type
$$
I_\eta^{(p)} = \int d \om \ \om^{n-3} \ n^{[\eta]} (\om)\ ({\om\over m})^p
\eqno(33)
$$
where $m$ is a mass scale typical of the external momenta components
$(p_0,{\vec p}),\ (q_0,{\vec q})$... (all external variables are supposed to
have comparable sizes). This can be re-expressed with the variable $z=\om/T$ as
$$
I_\eta^{(p)} =T^2\ ({T\over m})^p\ T^{-2\eps} \ \int^\infty_0 dz
\ {z^{1+p-2\eps}\over e^z-\eta} \eqno(34)
$$
where $n=4-2\eps$. These integrals are easily expressed in terms of $\G$ and
Riemann $\ze$ functions and we find:
$$
\eqalign{
I_{ 1}^{(p)} & = T^2 \ ({T\over m})^p \ T^{-\eps}\ \G(2+p-2\eps)
\ \ze(2+p-2\eps)\cr
I_{-1}^{(p)} & = (1-2^{-1-p+2\eps})\ I_{ 1}^{(p)} \cr}.\eqno(35)
$$
Defining ``soft" momenta as those with all components of ${\cal O}(eT),\
e\ll1$, then a Green's function with soft external momenta behaves as
$$
I^{(P)}_\eta \sim T^2\ ({1\over e})^p.
\eqno(35a)
$$
Clearly the dominant term comes from the largest value of $p,$ \ie from terms
in the integrand with the highest power of $L$, the loop momentum. In other
words, the dominant behavior arises from the parts in the integrand which are
leading when $l_i\sim T$, hence the name of hard loop approximation. This is
the basis of the resummation approach of Braaten and Pisarski [\cite{2,3}]
which consists in taking account of all hard loops at any order of perturbation
theory. Several examples are discussed below.

\vskip .50cm

Consider the massless fermion self-energy in the Feynman gauge (see fig.~6).
The general expression is (we simplify the notation by using $\Sigma_\a$
instead
of $\Sigma_{\a\a}$)
$$
\eqalign{
-i\Sigma_\a(P) & = +e^2 \int {d^nL\over(2\pi)^n} \ \ga_\nu \ (\slP+\slL)\
\ga^\nu  \Big[({1\over2}+n^B(l_0))\ (\D_R(L)-\D_A(L))\ \D_\a (P+L) \cr
&\ \ \ \ + ({1\over2}-n^F(p_0+l_0))\ (\D_R(P+L)-\D_A(P+L))
\ \D_{\bar\a}(L)\Big]. \cr}
\eqno(36)
$$
For a soft fermion $(p\sim eT)$, we can, according to the above discussion,
neglect $\slP$ compared to $\slL$ and with a change of variable in the second
term $(P+L\to-L)$ we arrive at
$$
-i\Sigma_\a(P)=-2(1-\eps)e^2 \int{d^{n-1}L\over(2\pi)^{n-1}}
\ {n^B(\om)+n^F(\om)\over2\om} [\slL\ \D_\a(P+L)+\slL'\ \D_\a(P+L')] \eqno(37)
$$
with $L=(\om,\vec l),\ L'=(-\om,\vec l)$. Consider now $\D_\a(P+L)$ for
example,
$$
\D_\a(P+L)={i\over2|\vec p+\vec l|}
\Bigg[{1\over p^0+\om-|\vec p+\vec l|+i\veps_\a} -
{1\over p^0+\om+|\vec p+\vec l|+i\veps_\a} \Bigg] \eqno(38)
$$
(we use the convention $\veps_R=\veps$ and $\veps_A=-\veps,\ \veps>0$).
Under the assumption that $P$ is soft, we can safely neglect the second term,
which behaves as $1/\om^2$, and keep only the first one which for large $\om$
reduces to
$$
\D_\a(P+L)={1\over2\om}\ {i\over p_0-\vec p\ \hat l + i \veps_\a},
\ \  \hat l={\vec l\over\om}. \eqno(39)
$$
In the hard loop approximation, only one of the two poles of the R/A propagator
is relevant, namely the pole associated to Landau damping [\cite{3}]. The pole
associated to particle production which is the only one at 0 temperature, is
suppressed by powers of the coupling constant (or equivalently by factors in
$1/T$). The fermion self-energy then takes the simple form
$$
\Sigma_\a(P)= {(1-\eps)e^2\over2\pi^2} \int d \om\ \om^{1-2\eps}
\ (n^B(\om) +n^F(\om)) \int {1\over2} {d\hat L\over(2\pi)^{1-2\eps}}
\ {\hat{\slL}\over P\hat L+i\veps_\a} \eqno(40)
$$
where we have introduced the light-like vector $\hat L = (1,\hat l)$. The
dimensional and the angular part of the loop integration factorize, as is
well-known [\cite{3,4}] and we can write our final result:
$$
\Sigma_\a(P) = m^2_{th} (\eps) \int {1\over2} {d\hat L\over(2\pi)^{1-2\eps}}
\ {\hat{\slL} \over P\hat L +i\veps_\a}. \eqno(41)
$$
In the 4-dimensional limit, the mass term $m^2_{th}(\eps)$ reduces to the
thermal mass squared [\cite{22}]
$$
m^2_{th} = {e^2 T^2\over 8} \eqno(42)
$$
a well-known result. Although not necessary here, we have worked in
$n$-dimensions for later purposes. We immediately have the useful property
$$
\Sigma_\a(-P) =-\Sigma_{\bar\a} (P). \eqno(43)
$$
The effective propagator is defined as
$$
\eqalign{
^*S_\a (P)  & = {i\over\slP-\Sigma_\a(P)} \cr
& = i {(\slP-\Sigma_\a(P))\over D^2_\a(P)} \cr} \eqno(44)
$$
where, in the second form, $D^2_\a(P)$ is a scalar which need not be specified
at this point. If $P$ is soft, of ${\cal O}(eT)$, eq.~(41) immediately shows
that $\Sigma_\a(P)$ is also of ${\cal O}(eT)$, hence the necessity of using the
resummed propagator eq.~(44) rather than the bare one for a consistent
calculation. If on the other hand $P$ is hard, of ${\cal O}(T)$, it appears
that $\Sigma_\a(P)$ is of ${\cal O}(e^2T)$ from eq.~(41) and therefore the self
energy gives a (negligible) correction of ${\cal O}(e^2)$ to the bare
propagator. In general, hard propagators need not be resummed in a leading
order calculation [\cite{3}]. We recall the antisymmetry property of the
effective fermion propagator which is a consequence of eq.~(43)
$$
^*S_\a(-P)=- ^*S_{\bar\a}(P). \eqno(44a)
$$
It will be used later to simplify calculations.

We could calculate, using the same technique, the vacuum polarization diagram
and obtain the scalar $\pi^T_\a(Q),\ \pi_\a^L(Q)$, \ie the transverse and
longitudinal polarization functions [\cite{23}] which are the analytic
continuations $\pi^L(q_0+i\veps_\a,\vec q)$, $\pi^T(q_0+i\veps_\a,\vec q)$ of
the imaginary time approach. Since we do not need these results in the
following we do not go into details and turn now to the three-point function.

\vskip .5cm

For the sake of completeness, we consider temporarily the case of massive
fermions and work in a general gauge where the photon propagators is denoted
${\cal P}^{\nu \rho}(L_1)\ \D_\a(L_1)$. We also consider the complete
expression not assuming the hard loop approximation.
It is easy to prove (in fig.~4 we assume $Q$ to be the photon momentum and $P$
the incoming fermion momentum):
$$
\eqalign{
V^\mu_{\a\be\de}(P,Q,R) =-e^2\int & {d^nL_1\over(2\pi)^n} \ \ga_\nu\ (\slL_2+M)
\ \ga_\mu\ (\slL_1+M)\ \ga_\rho {\cal P}^{\nu \rho}(L_1)\cr
&\Bigg[ ({1\over2}+n^B(l_{10}))\ \veps (l_{10})\ \de(L^2_1)
\ \D_\a(L_2)\ \D_{\bar\de}(L_3)    \cr
& +  ({1\over2}-n^F(l_{20}))\ \veps (l_{20})\ \de (L_2^2-M^2)
\ \D_\be (L_3)\ \D_{\bar\a}(L_1)  \cr
& +  ({1\over2}-n^F(l_{30}))\ \veps (l_{30})\ \de (L_3^2-M^2)
\ \D_\de(L_1)\ \D_{\bar\be}(L_2)\Bigg].  \cr} \eqno(45)
$$
In order to prove that the usual Ward identity holds true at finite temperature
we construct the scalar $Q_\mu V^\mu_{\a\be\de}(P,Q,R)$. Making use of the
following identities:
$$
\eqalign{
(\slL_2+M) \ \slQ\ (\slL_3+M) &\ \de(L_2^2-M^2)\ \D_\be(L_3) = i (\slL_2+M)
\ \de(L^2_2-M^2)    \cr
(\slL_2+M) \ \slQ\ (\slL_3+M) & \ \de(L_3^2-M^2)\ \D_{\bar\be}(L_2)=-i
(\slL_3+M)
\ \de(L_3^2-M^2)    \cr
(\slL_2+M) \ \slQ\ (\slL_3+M) & \ \de(L_1^2)\ \D_\a(L_2)\ \D_{\bar\de}(L_3) \cr
&=i\left((\slL_2+M)\ \D_\a(L_2)-(\slL_3+M)\
\D_{\bar\de}(L_3)\right)\de(L_1^2)\cr}
\eqno(46)
$$
we easily derive the following identity
$$
Q_\mu  V^\mu_{\a\be\de} (P,Q,R) = \Sigma_\a(P) - \Sigma_{\bar\de}(-R).
\eqno(47)
$$
This holds true not only for the $e^2T^2$ terms but also for the $e^2T$ and of
course the constant $(T=0)$ pieces. In the derivation, the thermal factors do
not play any particular role. Knowing that, at $T=0$, eq.~(47) is satisfied to
any loop order it can presumably be proven, by similar methods, that it is also
true at finite temperature [\cite{24}].

Let us now turn to the hard loop approximation. As is the case of the
self-energy it is found that only the Landau damping contribution has to be
considered. We are then justified in writing the propagators
$$
\eqalign{
\D_\a(L+P) & = {1\over2\om}\  {i\over P\hat L+i\veps_\a}\cr
\D_{\bar\de}(L-R) & = {1\over2\om}\  {-i\over R\hat L+i\veps_\de}, \cr}
\eqno(48)
$$
and recombining terms proportional to $({1\over2}-n_F)$ in eq.~(46) we have
(massless theory, Feynman gauge)
$$
\eqalign{
V^\mu_{\a\be\de}(P,Q,R) & =-{1\over2}(1-\eps)e^2\int {d^nL\over(2\pi)^{n-1}}
\hat{\slL}\ \ga^\mu\ \hat{\slL}\ (n^B(l_0)+n^F(l_0))\ \veps(l_0)\ \de(L^2) \cr
&\qquad \qquad \qquad \qquad {1\over P\hat L+i\veps_\a}\
{1\over R\hat L+i\veps_\de}	\cr
& = - m^2_{th}(\eps)\int{1\over2} {d\hat L\over(2\pi)^{1-2\eps}}
\ {\hat L^\mu\hat{\slL}
\over(P\hat L+i\veps_\a)(R\hat L+i\veps_\de)} \cr}
\eqno(49)
$$
The interesting feature about this formula is its independence on the retarded
or advanced prescription on the photon momentum $Q$: the analytic structure of
the vertex is entirely given by the prescription on the fermion momenta. As a
consequence, in the hard loop approximation we can write
$$
V^\mu_{\a R \de}(P,Q,R) - V^\mu_{\a A\de} (P,Q,R) =0 \eqno(50)
$$
(not all indices being equal in the above expression). Mathematically this
relation expresses the fact that the function $V^\mu_{\a\be\de}(P,Q,R)$ of
three
variables $P,Q,R$, considered to be independent has no singularity when
crossing the real $q_0$ axis. In other words
$$
{\rm Disc}_Q \ V^\mu_{\a R\de} (P,Q,R) =0 \eqno(50a)
$$
The independence of the variables $P,Q,R$ is understood in the following sense:
the retarded or advanced prescriptions on the momenta $P$ and $R$ respectively
are not dependent on that of $Q$ since $\be$ takes the values $R,A$ without
changing the indices $\a$ and $\de$. In the diagrammatic decomposition of
fig.~5, eq.~(49) is entirely represented by term a) with only the internal
photon line being cut
and the $i\veps$ prescription entirely carried by the internal fermion lines.
This property results from a rearrangement of terms in the integrand and, as a
consequence, the statistical weight attached to the cut photon line is not only
$2\pi({1\over2}+n^B(l_0))\ \eps(l_0)\ \de(L^2)$ as in the general case but
rather
$2\pi(n^B(l_0)+n^F(l_0))\ \eps(l_0)\ \de(L^2)$ for the hard loop case. Since
the
angular dependence factorizes out in the integrand we can perform the
integration over the energy and the length of the internal momentum variable
and thus recover the thermal mass factor. This property will simplify the
cut-structure of higher loop diagrams as will be seen later.

The contraction of the vertex function with $Q_\mu$ immediately yields
eq.~(47),
which now becomes a relation between Green's function evaluated in the hard
loop approximation. Pseudo Ward identities are also obtained [\cite{4}]
$$
\eqalign{
P\cdot V_{\a\be\de}(P,Q,R) & = - \Sigma_\de(R)\cr
R\cdot V_{\a\be\de}(P,Q,R) & = - \Sigma_\a(P) \cr} \eqno(51)
$$
which are only true at the leading $eT$ level.
We define now an effective vertex
$$
^*V^\mu_{\a\be\de} (P,Q,R)=\ga^\mu +V^\mu_{\a\be\de}(P,Q,R). \eqno(52)
$$
A dimensional analysis of eq.~(49) shows that if all external vertex momenta
are soft, then the function $V_{\a\be\de}^\mu$ is of ${\cal O}(1)$ like the
bare
vertex and, in that case, the effective vertex $^*V^\mu_{\a\be\de}$ has to be
used for a consistent calculation. If, on the contrary, one (and therefore
at least two)
external momenta are hard then the loop correction is down by at least a factor
$e^2T$ compared to the tree vertex.

\vskip .5cm

We turn now to the 4-point function and, as an example, we consider the case
with 2 external photons and an fermion-antifermion pair. We restrict
ourselves, for the moment, to
functions of type $C^{\mu\nu}_{R R R A}(P,Q_1,Q_2,R)$ and cyclic
permutations on the R/A indices. They are the Fourier transforms of the
retarded products of fields [\cite{14}]. Defining the sum of diagrams of
fig.~7 as $ie^2\ C^{\mu\nu}_{\a\be\ga\de}(P,Q_1,Q_2,R)$ we derive the hard loop
expression (the superscripts $\mu$ and $\nu$ refer to the Dirac indices),
$$
\eqalign{
C^{\mu \nu}_{\a\be\ga\de}(P,Q_1,Q_2,R) &= m^2_{th}(\eps) \int{1\over2}
{d\hat L\over(2\pi)^{1-2\eps}}\ \hat L^\mu\ \hat L^\nu\ \hat{\slL}
\ {1\over P\hat L+i\veps_\a}\ {1\over R\hat L+i\veps_\de}\cr
&\ \ \ \ \left({1\over(P+Q_1)\hat L+i(\veps_\a+\veps_\be)} +
        {1\over(P+Q_2)\hat L+i(\veps_\a+\veps_\de)}\right) \cr}
\eqno(53)
$$
when the sign of the $i\veps$ terms is entirely determined by the prescriptions
on the external legs. The Ward identities are easily obtained (we recall that
one index is equal to $A$ and all the others are equal to $R$)
$$
\eqalign{
Q_{1\mu} C^{\mu\nu}_{\a\be\ga\de}(P,Q_1,Q_2,R) = V^\nu_{\de\cdot(\a+\be)}
(R,Q_2,-R-Q_2)\ -\ V^\nu_{\a\cdot(\be+\de)}(P,Q_2,-P-Q_2) \cr}
\eqno(54)
$$
where the symbol $\cdot$ in the indices means that the corresponding R/A index
need not be specified.

	We turn now to the case of $C^{\mu\nu}_{A A A R }(P,Q_1,Q_2,R)$ where
we consider also the cyclic permutations on the indices. For this purpose it is
useful to introduce the normalized functions
$\widetilde C^{\mu\nu}_{A A A R}(P,Q_1,Q_2,R)$ defined by [\cite {18}],
[\cite {20}]
$$
\eqalign{
C^{\mu\nu}_{\a \be \ga \de}(P,Q_1,Q_2,R)= & {(n^F(p_0))^{\de_{\a A}}
(n^B(q_{10}))^{\de_{\be A}} (n^B(q_{20}))^{\de_{\ga A}} (n^F(r_0))^{\de_{\de
A}}
\over n(\de_{\a A}p_0+\de_{\be A}q_{10}+\de_{\ga A}q_{20}+\de_{\de A}r_0)} \cr
& \widetilde C^{\mu\nu}_{\a \be \ga \de}(P,Q_1,Q_2,R) \cr}
\eqno(55)
$$
which in the hard loop approximation is given by eq.~(53). This can be seen
either by direct calculation or by using the general relation [\cite{18}]
(only one index $\a,\ \be,\ \ga$ or $\de$ is equal to R)
$$
\widetilde C^{\mu\nu}_{\a \be \ga \de}(P,Q_1,Q_2,R)=
\big( C^{\mu\nu}_{\bar \a \bar \be \bar \ga \bar \de}(P,Q_1,Q_2,R) \big)^*
\eqno(56)
$$
The function $\widetilde C^{\mu\nu}_{\a \be \ga \de}(P,Q_1,Q_2,R)$ therefore
satisfies eq.~(53) which holds true then whenever three of
the indices are identical. We can go back to the full Green functions and we
get, for example,
$$
\eqalign{
Q_{1\mu} C^{\mu\nu}_{R R R A}(P,Q_1,Q_2,R)  = & \ \Gamma^\nu_{A R R}
(R,Q_2,-R-Q_2)\ -\ \Gamma^\nu_{R R A}(P,Q_2,-P-Q_2) \cr
Q_{1\mu} C^{\mu\nu}_{A A A R}(P,Q_1,Q_2,R)  = &
-\ {n^F(p_0)\ n^B(q_{10})\over n^F(p_0+q_{10})}\ \Gamma^\nu_{R A A}
(R,Q_2,-R-Q_2)   \cr
&  +\ {n^B(q_{10})\ n^F(p_0+q_{20})\over n^F(p_0+q_{10}+q_{20})}\
\Gamma ^\nu_{A A R}(P,Q_2,-P-Q_2) \cr}
\eqno(57)
$$
We note that in the second case there appears a pre-factor in front of the
vertex function depending explicitely on the photon momentum $Q_1$: this is
necessary since the 4-point function carries such factors whereas in the
3-point
functions momentum $Q_1$ does not appear. This suggests that, in general, Ward
identities take a simpler form in terms of the normalized functions of type
$V^\mu_{\a \be \ga}$ and $\widetilde C^{\mu \nu}_{\a \be \ga \de}$, which
are analytic continuations of the imaginary time formalism, than in terms of
the full $R/A$ functions. This can be verified when considering the more
complicated case of $C^{\mu\nu}_{A R A R}(P,Q_1,Q_2,R)$ which in the hard loop
approximation reduces to
$$
\eqalign{
& C^{\mu \nu}_{A R A R}(P,Q_1,Q_2,R) = m^2_{th}(\eps) \int{1\over2}
{d\hat L\over(2\pi)^{1-2\eps}}\ \hat L^\mu\ \hat L^\nu\ \hat{\slL}
\ {1\over P\hat L-i\veps}\ {1\over R\hat L+i\veps}\cr
&\ \ \left({1-n^F(p_0)-n^F(r_0+q_{20}) \over (P+Q_1)\hat L+i\veps} +
           {1+n^B(q_{20})-n^F(-r_0-q_{20}) \over (P+Q_1)\hat L-i\veps} +
           {1-n^F(p_0)+n^B(q_{20}) \over(P+Q_2)\hat L-i\veps}\right) \cr}
\eqno(58)
$$
It leads to the following Ward identity:
$$
\eqalign{
Q_{1\mu} C^{\mu\nu}_{A R A R}(P,Q_1,Q_2,R)  = &\ \Gamma^\nu_{A A R}
(P,Q_2,-P-Q_2)\ -\ \Gamma^\nu_{R A A}(R,Q_2,-R-Q_2) \cr
& \ + {n^F(p_0)\ n^F(r_0+q_{20})\over n^F(-q_{10})}\ \Gamma^\nu_{R A A}
(R,Q_2,-R-Q_2)   \cr }
\eqno(59)
$$
The last two terms appear because the prescription on the energy variable
$r_0+q_{20}$ is not defined by the external conditions and in such a case
the R/A formalism selects a particular linear combination of the retarded
and advanced continuations.

\vskip .50cm

\noindent

{\bf V. 2-point function in the multi-loop approximation}

\vskip .50cm

In the framework of the Braaten-Pisarski resummation we need the expression of
Green's functions beyond the one-loop approximation. In particular, one is led
to evaluate two-point functions with vertex corrections such as shown in
Fig.~8a. We derive a general expression for the two-point function at the
multi-loop level, independently of the hard loop approximation to which we
return at the end of the section.

In the case of interactions involving only three particle vertices, it is
always possible to cut the diagram in such a way as to have only two particle
intermediate states (Fig.~8b). This allows us to express the self-energy in
terms of the vertex functions $\G_{\a\be\de}$
$$
-i\G^{(j+k+1)}_{\be\be'}(Q)=-e^2\int{d^nP\over(2\pi)^n}\ \D_\a (P)
\ \G^{(j)}_{\a\be\bar\de}(P,Q,-R)\ \D_\de(R)\
  \G^{(k)}_{\bar\a\bar\be ' \de}(-P,-Q,R)
\eqno(60)
$$
where the various $\Gamma$ functions carry the superscripts
$(j),\ (k)$, and $(j+k+1)$ to denote the number of loops at which they have
been respectively evaluated.
Introducing the normalized functions as in eq.~(28) and regrouping terms we
obtain (dropping irrelevant spin factors)
$$
\eqalign{
& -i  \G^{(j+k+1)}_{\be\be}(Q) =-e^2 \int {d^nP\over(2\pi)^n} \cr
& \Bigg\{ \left({1\over2}+\eta_P n^{[\eta_P]}(p_0)\right)
{\rm Disc}_P \left[\D_R (P)\ V^{(j)}_{R\be A} (P,Q,-R)
\D_R(R)\ V^{(k)}_{A\bar\be R} (-P,-Q,R)\right] \cr
& +\left({1\over2}+\eta_Rn^{[\eta_R]}(r_0)\right)
{\rm Disc}_R \left[\D_A (P)\ V^{(j)}_{A\be A} (P,Q,-R)
\D_R(R) V^{(k)}_{A\bar\be R} (-P,-Q,R)\right]\Bigg\} \cr}
\eqno(61)
$$
where the symbol ${\rm Disc}_P$ applied on a function $F_{\a\be\de}(P,Q,R)$
means taking the discontinuity in the energy
variable $p_0$ of the function as defined by
$$
{\rm Disc}_P\ F_{R\be\de}(P,Q,R)= F_{R\be\de}(P,Q,R) -F_{A\be\de}(P,Q,R)
\eqno(62)
$$
(see the discussion around eq.~(50)). The ingredients to derive this
relation are the same as those to obtain the one loop expression, namely the
diagonality of propagators, the validity of eq.~(27) for the dressed vertex and
the property that terms with poles only on one side of the real axis of the
loop energy variables give a vanishing contribution when no statistical weight,
depending on the loop energy variable, is attached to them. Despite a rather
cumbersome notation, the structure of eq.~(60) is rather simple and similar to
the one-loop result. Consider, as an example, the calculation of $\G_{RR}(Q)$.
The function in the first line, whose ${\rm Disc}_P$ should be evaluated, is
trivially  obtained with propagators and vertex functions in the retarded $P$
and $Q$ momenta (recall also the crossing relation eq.~(13)). Momentum $R$ is
necessarily retarded because of momentum conservation. Likewise the second line
is constructed from the retarded momenta $Q$ and $R$. However, $P$ is now of
the advanced type since, using $p_0=r_0-q_0$ and recalling that taking the
discontinuity in $r_0$ puts $r_0$ as the real axis, the prescription for $p_0$
is that of $-q_0$, \ie advanced. Taking the discontinuity of these products of
propagators and vertices generates two types of terms:
${\rm Disc}_P\ \D_R(P)= 2\pi\ \veps(p_0)\ \de(P^2-M^2)$ puts the $P$ line on
shell while
${\rm Disc}_P\ V_{R\be\de}(P,Q,-R)$ amounts to taking the cut contribution of
the vertex.

Likewise it is easy to prove
$$
\G_{\be\bar\be}(Q)=0.
$$

As an application we may consider the case when $V_{\a\be\de}$ consists in a
one loop diagram and study ${\rm Disc}_P\ V_{\a\be\de}(P,Q,R)$. We have already
seen that the vertex is expressed as a sum of the diagrams (eq.~(29) and
fig.~5). Taking the discontinuity with respect to $p_0$ means, according to
eq.~(62), putting the internal momentum lines carrying the index $\a$ on shell.
An examination of eq.~(29), leads then to taking the line $L_2=L_1+P$ on shell
when $L_1$ is cut and $L_1=L_2-P$ on shell when $L_2$ is cut. This we symbolize
by a cross on the corresponding line (see fig.~9). We use a different symbol
to denote this
cut because the statistical weight associated to taking ${\rm Disc}_P$ is
$({1\over2}+\eta_P n^{[n_P]}(p_0))$ rather that the thermal factor appropriate
for the internal line. Therefore, ${\rm Disc}_P\ V_{\a\be\de}$ in eq.~(61)
picks up the 2-particle cut contribution in $p_0$ while
${\rm Disc}_P\ \D_R(P)$ selects the pole in $p_0$. One can easily convince
one-self, by such arguments that the self-energy in the multi-loop
approximation retains a tree structure as it does at one loop.

In the next section we will be led to consider the imaginary part of the two
point functions which is evaluated by constructing $\G_{RR}(Q)-\G_{AA}(Q)$. It
can be written as
$$
\eqalign{
\G_{RR}(Q) & -\G_{AA}(Q)  =-ie^2\int {d^nP\over(2\pi)^n}
\Bigg\{ \left({1\over2}+\eta_P n^{[\eta_P]}(p_0)\right) \cr
&\qquad \Bigg[{\rm Disc}_P \D_R(P) V_{RRA}(P,Q,-R)\D_R(R) V_{AAR}(-P,-Q,R) \cr
&\qquad -{\rm Disc}_P\D_R(P)V_{RAR}(P,Q,-R)\D_A(R)V_{ARA}(-P,-Q,R)\Bigg]\cr
&\qquad +({1\over2}+\eta_Rn^{[\eta_R]}(r_0)) \cr
&\qquad \Big[{\rm Disc}_R\ \D_A(P) V_{ARA}(P,Q,-R)\D_R(R) V_{RAR} (-P,-Q,R)\cr
&\qquad {-\rm Disc}_R \D_R(P) V_{RAA} ((P,Q,-R)\D_R(R) V_{ARR} (-P,-Q,R)\Big]
\Bigg\}. \cr}
\eqno(63)
$$
The difference of the ${\rm Disc}_P$ or ${\rm Disc}_R$ expressions above have
a simple interpretation. Let us remark that in the first square brackets the
two terms differ by the indices associated to momenta $Q$ and $R$ while in the
second ones the terms differ by the indices associated to momenta $Q$ and $P$.
Consider the ${\rm Disc}_P$ case. As we have seen, taking the discontinuity
with respect to the variable $p_0$ amount to putting the energy $p_0$ on the
real axis. By ``momentum conservation", $r_0=-p_0-q_0$, the retarded/advanced
prescription on the internal momentum $R$ is now entirely determined by that on
the external momentum $Q$: $r_0$, in that sense, becomes a function of $q_0$.
This is expressed by the fact that the ${\rm Disc}_P$ terms differ by their
indices in both $Q$ and $R$. Keeping this in mind, the square brackets then
isolate the discontinuity in $q_0$ of the considered functions and we may write
$$
\eqalign{
{\rm Disc}_Q\ {\rm Disc}_P\ \D_R(P) & \ V_{RRA} (P,Q,-R)\ \D_R(R)
\ V_{AAR}(-P,-Q,R)    \cr
&={\rm Disc}_P\ \D_R(P)\ V_{RRA}(P,Q,-R)\ \D_R(R)V_{AAR}(-P,-Q,R) \cr
&-{\rm Disc}_P\ \D_R(P)\ V_{RAR}(P,Q,-R)\ \D_A(R)V_{ARA}(-P,-Q,R)\cr}
\eqno(64)
$$
where $R$ is considered as a function of $Q$, when ${\rm Disc}_P$ is evaluated.

If we turn now to the hard loop approximation we find that the singularity
structure of the effective vertices simplifies considerably. A case in point is
the QED vertex studied in the previous section where it was shown that (in the
hard loop approximation) ${\rm Disc}_Q\ V_{\a R\de}(P,Q,R)=0$ (eq.~(50a)). This
function enters the calculation of the vacuum polarization tensor
$\pi_\be^{\mu\nu}(Q)$ to be considered shortly. In eq.~(64), as a consequence,
we can write ${\rm Disc}_R\ {\rm Disc}_P$ for the first term and
${\rm Disc}_P\ {\rm Disc}_R$ for the second one. Furthermore, from eq.~(49) it
appears that the dependence on $P$ and $R$ factorizes in the integrand which we
can write as a product of two functions $f^{(1)}_\a(P)\ f^{(2)}_\de(R)$: the
double discontinuity becomes then a product of discontinuities so that
$$
\eqalign{
\Gamma_{RR} (Q)-\Gamma_{AA}(Q) &=-ie^2\int{d^nP \over(2\pi)^n}
\Bigg\{ (1-n^F(p_0))
\ {\rm Disc}_P f^{(1)}_R(P) \ {\rm Disc}_R f^{(2)}_R (R)    \cr
&\qquad\qquad\qquad\qquad + ((1-n^F(r_0)) \ {\rm Disc}_P f^{(1)}_A(P)
\ {\rm Disc}_R f^{(2)}_R(R)\Bigg\} \cr}
\eqno(65)
$$
where we have considered the case with an internal fermion loop.
Combining both terms and using the  detailed balance relation
$$
n^F(r_0)-n^F(p_0)=(1-e^{\be q_0})n^F(r_0)n^F(-p_0)
$$
it comes out
$$
\eqalign{
\Gamma_{RR}(Q)-\Gamma_{AA}(Q) =- ie^2\int & {d^{n-1}P\over(2\pi)^{n-1}}
(1-e^{\be q_0}) \int dp_0dr_0\ n^F(r_0) n^F(-p_0)\ \de(q_0+p_0-r_0)\cr
&{1\over2\pi} {\rm Disc}_P\ f^{(1)}_R(P) \ {\rm Disc}_R\ f^{(2)}_R(R). \cr}
\eqno(66)
$$
This is to be compared to the formula in [\cite{5}], [\cite{6}] which expresses
the discontinuity of the two-point function as an integral over real energies
$$
\eqalign{
& \int{d^{n-1}P\over(2\pi)^{n-1}} {\rm Disc}\ T  \sum_{p_0} f^{(1)}(p_0,\vec p)
 f^{(2)} (q^0-p^0,\vec q-\vec p)\cr
&=i \int {d^{n-1}P\over(2\pi)^{n-1}}(1-e^{\beta q_0}) \int d\om d \om'
n^F(\om')n^F(\om') \de
(q_0-\om-\om')2\pi\rho_1(\om,\vec p)\rho_2 (\om',\vec q-\vec p)\cr}
\eqno(67)
$$
where the spectral density $\rho_i$ is defined as, for example,
$\rho_i(\om,\vec p)=2\pi {\rm Disc}\ f^{(i)}(p_0,\vec p)$.

The structure of eqs.~(61), (63) is not changed, if instead of the bare
propagators $\D_\a(P),\ \D_\de(R)$ we use in these equations the effective
propagators $^*\D_\a(P)$, $^*\D_\de(R)$. The poles near the real axis are
shifted away from it and become singularities in the same half-plane of the
loop energy variable. In particular, the crucial ingredient in deriving these
equations, namely that an advanced propagator has no pole in the lower
half-plane and a retarded one has no pole in the upper half-plane,
is not affected. When evaluating the discontinuities of these
effective propagators we will not only get the pole contributions as in the
bare
case but also two particle cuts associated to the Landau damping mechanism.

\vskip .50cm
\noindent
{\bf VI. Real soft photon production in a quark-gluon plasma.}

\vskip .50cm

The problem of photon emission in a quark-gluon plasma has already been
considered for several cases: soft virtual photon at rest [\cite{5}] or moving
[\cite{6}],[\cite{25}]  and hard real photon [\cite{7}],[\cite{8}]. We apply
the above
formalism to soft real photon [\cite{26}]. We assume massless quarks and we
introduce the strong interactions coupling, denoted $g$, assuming
$g\ll1$. The photon production rate [\cite{27}], assuming $q_0$ positive for
simplicity,
$$
q_0 {d \si \over d^3 q} = - {1 \over (2 \pi)^3} n_B (q_0)
\ {\rm Im} \Pi^\mu_\mu (Q) |_{Retarded}
\eqno(68)
$$
is related to the trace of the polarization tensor which in the following we
denote for short $\Pi_R (Q)$, after summing over the photon polarization
states.

Consider first the one-loop contribution to ${\rm Im} \ \Pi_R(Q)$
before resummation,
i.e. using bare propagators and vertices. Using eq.~(18), it is trivially found
that ${\rm Im} \ \Pi_R(Q)$ vanishes (but ${\rm Re} \ \Pi_R(Q) \neq 0$).
The reason for this
is simple: the only kinematical configurations possibly contributing to
${\rm Im}\ \Pi_R$ are the collinear decay of the photon into a
$q \bar{q}$ pair or the
collinear emission or absorbtion of the photon by a quark or an antiquark in
the plasma. However helicity conservation at the $\ga q \bar{q}$ vertex forbids
these processes.

We turn now to the effective theory and evaluate the same "one loop" diagram
using effective propagators and vertices as shown in Fig. 10 a). The tadpole
diagram (Fig. 10 b) vanishes
because it is traceless [\cite{5}]. In the framework of the resummed
perturbative series, the self-energy like diagram gives the dominant
contribution to the production of soft virtual photons. We have
$$
\eqalign{
-i \Pi_R(Q) &= - e^2 \int {d^nP \over (2 \pi)^n} (1-2\ n^F(p_0))\ g_{\mu \nu}
\cr
& \ \ \ {\rm Disc}_P {\rm Tr} \Big[ \ ^* \cS_R(P)^* V^\mu_{RRA} (P,Q,-R)
^*\cS_R(R)^*V^\nu_{RRA}(P,Q,-R) \Big] \cr}
\eqno(69)
$$

Some words of explanation are required concerning this equation. The effective
propagator $^*\cS_\a$ and vertex $^*V^\mu_{\a\be\de}$ are understood to contain
QCD corrections and not QED corrections. At the order at which we do the
calculation it simply amounts to substituting $e^2 \rightarrow C_F g^2$ in
eq.~(42) to take into account the change in coupling as well as the color
factor. We thus define the thermal mass to be now
$$
m^2_{th} = C_F {g^2T^2 \over 8},
\eqno(70)
$$
all other equations in the previous section being unchanged. We have applied
the property of eq.~(31) to the effective vertex eq.~(52) so that the same
vertex function appears twice in the integrand above. Finally, the expected
term proportional to $({1 \over 2} - n_F(r_0))$ can be reduced to the one above
after the change of variable $r_0=-p_0$ and the use of eqs.~(31) and (43),
hence the factor $1-2\ n^F(p_0)$. Writing out the explicit form of
$^*V^\mu_{\a \be \de}$ we thus have to calculate the diagrams of Fig. 11. We
ignore
the first one which does not present any difficulty and turn to the second (or
equivalently the third) one which will be shown to present a collinear
divergence. We have:
$$
\eqalign{
-i \Pi_R(Q) |_b = -e^2 m^2_{th} (\eps) \int & {d^n P \over (2 \pi)^n} \int{1
\over 2} {d \hat{L} \over (2 \pi)^{1-2 \eps}} (1-2\ n^F (p_0) ) \cr
& {\rm Disc}_P {{\rm Tr}( ^*\cS_R(P)\ \hat{\slL}\ ^*\cS_R(R)\ \hat{\slL})
 \over (P \hat L + i \eps) ( R \hat L + i \eps)} \cr}
\eqno(71)
$$

We choose to carry the $\int dp_0$ integration by closing the contour in the
upper half-plane. Writing ${\rm Disc}_P$
explicitely as the difference of two terms,
$$
{\rm Disc}_P Tr  = {{\rm Tr}( ^*\cS_R(P)\ \hat{\slL}\ ^*\cS_R(R)\ \hat{\slL})
\over (P \hat L + i \eps) ( R \hat L + i \eps)}
- {{\rm Tr}( ^*\cS_A(P)\ \hat{\slL}\ ^*\cS_R(R)\ \hat{\slL}) \over
(P \hat L - i \eps) ( R \hat L + i \eps)}
\eqno(72)
$$
we see that no contribution arises from the first term since all its
singularities are located below the real axis. On the contrary, the second one
contains a pole in the upper half plane ($1 / (P \hat L - i \eps)$,
coming from the hard loop in the
effective vertex as well as a singularity from the resummed propagator
$^*\cS_A(P)$. Let us concentrate on the pole contribution. It comes out
$$
\eqalign{
-i \Pi_R(Q) |_b  = i e^2 m^2_{th} (\eps) \int & {d^n P \over (2 \pi)^{n-1}}
{1 \over 2} \int {d \hat L \over (2 \pi)^{1-2 \eps}}\ \de (P \hat L)\
(1-2n^F (p_0) ) \cr
& {{\rm Tr}(^*\cS_A(P)\ \hat{\slL}\ ^*\cS_R(R)\ \hat{\slL})
\over (Q \hat L + i \eps)} \cr}
\eqno(73)
$$
where the $p_0$ integration is understood now to run along the real axis. The
denominator $Q\hat{L} + i \eps$ is nothing but $R\hat{L} + i \eps$ where the
$\de$-function constraint has been used. Both $Q$ and $\hat{L}$ being
light-like this factor leads to a collinear divergence when the angular
integration $\int d \hat{L}$ is performed. From now on, we are only interested
in this diverging part neglecting all finite terms in the calculation. To
evaluate the residue at the pole, it is enough to set $\hat L = \hat Q =
Q/q$ in the above, except of course, in the denominator, leading to the rather
simple expression
$$
\eqalign{
-i \Pi_R (Q) |_{b,sing} = ie^2 m^2_{th}(\eps) \int & {d^n P \over (2
\pi)^{n-1}}
\ \de(P \hat Q)\ (1-2n^F (p_0)) {\rm Tr}(^*\cS_A(P)\ \hat{\slQ}\ ^*S_R(R)\
\hat{\slQ}) \cr
& {1 \over 2} \int {d \hat L \over (2 \pi)^{1-2\eps}} {1 \over Q \hat L
+ i \eps} \cr}
\eqno(74)
$$

The angular integration is understood in $n-2$ dimensions $(n=4-2 \eps)$ and
its real part has a pole in $\eps$. Therefore we define
$$
{\ga (\eps) \over \eps} = - {1 \over 2} \int {d \hat L \over (2 \pi)^{1-2\eps}}
{q \over Q \hat L + i \eps}
\eqno(75)
$$
neglecting a finite imaginary piece. The diverging contribution from the two
diagrams containing one hard loop effective vertex is therefore
$$
\eqalign{
\Pi_R(Q) |_{b+c, sing} = 2\ e^2\ {m^2_{th} (\eps) \over q} {\ga (\eps) \over
\eps} \int & {d^n P \over (2 \pi)^{n-1}} \de (P \hat Q) (1 -2n^F (p_0)) \cr
& {\rm Tr}( ^*S_A (P)\ \hat{\slQ}\ ^*S_R (R)\ \hat{\slQ})
\cr}
\eqno(76)
$$

To sum up, the diverging contribution arises from the momentum configuration
where both $P\hat{L}$ and $R\hat{L}$ vanish that is when the fermion
propagators coupling to the external photon both approach their mass-shell
condition. It is interesting and somewhat paradoxical that such a divergence is
a property of the resummed theory which tells us to use the effective
propagators $^*\D_\a(P)$ and $^*\D_\de(R)$ rather than the bare ones. Had we
used the latter, coming back to eq.~(71), we would have found that the trace
reduces to
$$
{\rm Tr} \slP \hat{\slL}  \slR \hat{\slL} = 8 P \hat L \ R \hat L
\eqno(77)
$$
cancelling both poles at the origin of the collinear divergence.

It is worth noting that the collinear divergence in eq.~(75) is related to the
vanishing mass of the photon. If $Q^2$ were slightly off-shell the pole in
eq.~(75) would never be reached in the time like case or would be defined
through a principal
value prescription in the space like case.

Let us turn now to the last diagram with two hard loop effective vertices. It
is explicitely
$$
\eqalign{
-i \Pi_R(Q) |_d  = & -e^2 m^4_{th}(\eps) \int {d^n P \over (2 \pi)^n}\ {1 \over
2} \int {d \hat L_1 \over (2 \pi)^{1-2 \eps}}\ {1 \over 2} \int {d \hat L_2
\over (2 \pi)^{1-2\eps}} \cr
& {\rm Disc}_P \Big( { \hat L_1  \hat L_2 \
{\rm  Tr}(^*\cS_R(P)\ \hat{\slL}_1\ ^*\cS_R(R)\ \hat{\slL}_2)
\over (P \hat L_1 + i \eps) (R \hat L_1 + i \eps) (P \hat L_2 + i \eps)
(R \hat L_2 + i \eps)} \Big) \cr}
\eqno(78)
$$
As in the previous case, a collinear divergence will arise from the collinear
configuration $\hat{L}_1 = \hat{Q}$, leading to a pole in $\eps$. It is to be
noted that the configuration $\hat{L}_2 = \hat{Q}$, together with the
collinearity
condition on $\hat{L}_1$ does not lead to a double pole because of the
$\hat{L}_1 \hat{L}_2$ factor. Setting thus $\hat{L}_1 = \hat{Q}$ in the
denominator we construct in fact a combination of type $Q_\mu
V^\mu_{\a\be\bar{\de}} (P,Q,-R)$ which can be immediately  reduced via the
Ward identity eq.~(47). More precisely we have in the integrand
$$
m^2_{th} (\eps) {1 \over 2} \int {d \hat L_2 \over (2 \pi)^{1-2 \eps}}
\ {\hat Q \hat{L}_2 \ \hat{\slL}_2 \over (P \hat L_2 + i \eps_\a) (R \hat L_2
 + i \eps_\de)} = {1 \over q} (\Si_\a (P) - \Si_\de (R)).
\eqno(79)
$$

Recasting the $\int dp_0$ integration as in eq.~(73) and exhibiting the
collinear divergence it comes out
$$
\eqalign{
-i \Pi_R |_{d, sing} = & -2\ i\ e^2\ m^2_{th}(\eps)\ {\ga (\eps) \over \eps}
\int {d^n
P \over (2 \pi)^{n-1}}\ \de (P \hat Q)\ (1-2\ n^F(p_0)) \cr
& \ \ {1 \over q}\ {\rm Tr}
\big( ^*S_A(P)\ \hat{\slQ}\ ^*S_R(R)\ (\Si_A(P) - \Si_R(R))\big) \cr}
\eqno(80)
$$

The factor 2 is introduced to account for the case when $\hat{L}_2$ is
collinear to $\hat{Q}$. Using the definition eq.~(44) of the effective
propagator the trace term is reduced to
$$
i\ \big({\rm Tr}(^*S_A (P) \hat{\slQ}) - {\rm Tr}(^*S_R(R) \hat{\slQ})\big)
- q\ {\rm Tr} (^* S_A (P)\ \hat{\slQ}\ ^*S_R(R)\ \hat{\slQ})
\eqno(81)
$$

It is immediately apparent that the last term exactly compensates the single
effective vertex pole contribution eq.~(76). We are thus left, adding all
pieces together, with
$$
\eqalign{
\Pi_R(Q) |_{sing} = i\ 2\ e^2\ {m^2_{th} (\eps) \over q^2}\ {\ga(\eps) \over
\eps}
\int & {d^n P \over (2 \pi)^{n-1}} \de (P \hat Q) (1-2n^F(p_0)) \cr
&\big({\rm Tr}(^*S_A(P)\ \hat{\slQ}) - {\rm Tr}(^* S_R(R)\ \hat{\slQ}) \big)
\cr} \eqno(82)
$$

The  $\Pi_A(Q)$ 2-point function is easily deduced from this equation by
replacing $^*S_\a(P)$ by $^*S_{\bar{\a}}(P)$ and the same for $^*S_\de(R)$. We
introduce now the usual parametrization of the effective fermion propagator
$$
^*S_R(P) = i \sum_{s=\pm1} {\hat P_s \over D^s_R (p_0 + i \eps, \vec p)}
\eqno(83)
$$
where $\hat{P}_s$ is the light-like vector $\hat{P}_s = (1,s\hat{p})$ and
$D^s(P)$ is a scalar function whose expression can be found in [\cite{5}],
[\cite{6}]. The sum runs over the two propagating modes of the thermalized
fermion. From the general property of 2-point functions, eq.~(21), and the
definition eq.~(44), we can write
$$
^*S_A(P) = - (^*S_R (P))^{c.c} = i \sum_s {\hat P_s \over (D^s_R (p_0 + i \eps
, \vec p))^*}
\eqno(84)
$$
which expresses the advanced effective propagator as the
complex-conjugate of the retarded one. It is customary to use the notation
$$
{1 \over D^s_{R,A}} = \a_s(P) \mp i \pi \be_s (P)
\eqno(85)
$$

We can now construct the imaginary part of the polarization function
$$
\eqalign{
{\rm Im} \Pi_R(Q) = 2\ e^2\ {m^2_{th} (\eps) \over q^2} {\ga(\eps) \over \eps}
\int
& {d^n P \over (2\pi)^{n-1}}\ \de (P \hat Q)\ (1-2n^F(p_0)) \cr
& \pi \sum_s \Big( \be_s (P){\rm Tr} \hat{\slP_s} \hat{\slQ}
+ \be_s(R) {\rm Tr} \hat{\slR_s}  \hat{\slQ} \Big) \cr}
\eqno(86)
$$

The traces are easily evaluated and give $4(1-s\hat{p} \hat{q})$ and $4(1-s
\hat{r}\hat{q})$ respectively, which  when taking into account the
$\de$-function constraint in the integrand reduced to $4(1-sp_0/p)$ and $4(1-s
r_0/r)$ leading to
$$
\eqalign{
{\rm Im} \Pi_R(Q) = 8\ e^2\ {m^2_{th} (\eps) \over q^2} {\ga(\eps) \over \eps}
\int
& {d^n P \over (2\pi)^{n-1}}\ \de (P \hat Q)\ (1-2n^F(p_0)) \cr
& \pi \sum_s \Big( (1-{sp_0 \over p}) \be_s (P) + (1- {sr_0 \over r}) \be_s (R)
\Big) \cr}
\eqno(87)
$$

If we remember that the $\be_s$ function are proportional to the same
$(1-sp_0/p)$ type factor with a positive coefficient we see that the integral
does not vanish and therefore the soft fermion loop contribution to the $\ga$
production rate in a plasma is divergent. It is interesting to note that the
condition $\de(P\hat{Q})$ enforces that $P$ (and also $R$ since $\de(R\hat{Q})
= \de(P\hat{Q}))$ be space like $(|p_0|<p), \ (|r_0| <r)$. The residue of the
collinear pole is entirely due to the Landau damping contribution to the
fermion effective propagator.

\vskip .50cm

An alternative derivation of eq.~(87) has recently been given in the imaginary
time formalism [\cite{26}]. It seems appropriate now to contrast the two
approaches. Since we work here in the real time formalism we can use the usual
Dirac algebra familiar from the $T=0$ case and we do not have the added
complication to redefine the algebra in Euclidean space. A more important
point derives from the fact that we calculate the full retarded (or advanced)
2-point function and not only its imaginary part. We manipulate 4-dimensional,
or rather $n-$dimensional integrals and the integrands, except for statistical
weights, keep a covariant form very similar to the $T=0$ case (see \eg
eqs.~(71) and (78)): because we work in $n-$dimensions we can evaluate the
integral over the energy by closing contours in the appropriate half-planes,
as we do in eq.~(71) and eq.~(78), to pick-up only the pole contributions which
lead to the singular behavior; since after this manipulation the integrand
keeps its covariant form we easily use the Ward identity (eq.~(47)) to extract
the final result. In contrast, in the ITF approach, the imaginary part of the
2-point function is directly calculated as an integral of spectral functions
over the real energy axis: the various pieces of the integrand have to be
decomposed into their principal values and $\de-$functions to extract the
divergent behavior and since the covariant form is lost in favor of the
spectral functions the Ward identity cannot be directly used to show the
partial compensation between the different terms. An explicit calculation of
all
terms is necessary to obtain the final results.

\vskip .50cm

\noindent
{\bf VII. Conclusions}

\vskip .50cm

	In this work, we have pursued the study of the R/A formalism and
derived, in particular, an important equation expressing the discontinuity of
the 2-point function at any loop order. We have also formulated, in this
approach, the hard loop expansion for the case of QED. We used the R/A
formalism to calculate the rate of real soft phton production in a QCD plasma
and found the result (eq.~(87)) to be divergent in
agreement with a recent calculation performed
in the imaginary time formalism [\cite{26}]. Compared to that work we find the
calculation in the $R/A$ formalism somewhat simpler, since, using contour
integration, we can extract the relevant contributions of poles in the complex
energy plane. We avoid distinguishing between principal part and $\de$-function
contributions of a pole as when the integration is performed on the real axis.
Also the cancellation of the single hard vertex loop diagrams with the double
hard vertex loop diagram is clearly
attributed to the thermal Ward identity eq.~(47). The structure of the
remaining diverging term involves the spectral function of the effective soft
fermion propagator. It is interesting to see how, in the framework of the
effective theory diagrams with different number of loops partially compensate
one another.

There are several ways the divergent result of eq.~(87) could be regularized.
In [\cite{26}] it was proposed to introduce a soft cut-off of order $gT$.
Another possibility is to include corrections to the hard internal fermion
lines which are the cause of the collinear divergence when they approach the
mass-shell condition. This is in the spirit of several attempts to define the
damping rate of a hard fermion. For example, including a width of $\cO (g^2T)$
on the hard internal fermion lines [\cite{28}] would displace the poles away
from the real axis and replace the $1/\eps$ divergence by $\ln(1/g)$. Finally,
we have considered in our calculation only the contribution of diagrams with
two internal soft fermion lines (see Fig.11). As discussed above the hard
fermion loop is kinematically allowed but it does not contribute at the lowest
order using bare propagators because of helicity conservation at the vertices.
Introducing  corrections to hard loops avoids this constraint. It would be
interesting to compare the order of such contributions to the result of
eq.~(87).

\vskip .50cm
\noindent
{\bf  Acknowledgements}

\vskip .50cm

We would like to thank R.~Baier, J.~Kapusta, R.~Kobes, S.~Peign\'e,
R.~Pisarski, D.~Schiff and A.~ Smilga for discussions. We also thank R.~Kobes
for a critical reading of the manuscript. One of us (PA) would like to thank
J.~Kapusta and E.~Shuryak for a very enjoyable stay at ITP. This research was
supported in part by NSF grant No. PHY89-04035.

\vfill\eject

\noindent
{\bf References}\break
\noindent

\item{[1]} 	See for example N.P. Landsman and Ch.G. van Weert,
		Phys. Rep. {\bf 145} (1987) 141;
\item{ }	J.I.~Kapusta, Finite Temperature Field Theory, Cambridge
		Monographs on Mathematical Physics, Cambridge University
		Press, 1989.
\item{[2]}	R.D. Pisarski, Phys. Rev. Lett. {\bf 63} (1989) 1129;
		Nucl. Phys. {\bf A498} (1989) 423c.
\item{[3]}	E. Braaten and R.D. Pisarski, Nucl. Phys. {\bf B337} (1990)
		569; Nucl. Phys. {\bf B339} (1990) 310.
\item{[4]}	J.~Frenkel and J.C.~Taylor, Nucl. Phys. {\bf B334} (1990)
		199;
\item{ }	J.C.~Taylor and S.M.H.~Wong, Nucl. Phys. {\bf B346} (1990) 115.
\item{[5]}	E. Braaten, R.D Pisarski and T.C. Yuan, Phys. Rev. Lett.
		{\bf 64} (1990) 2242.
\item{[6]}	S.M.H.~Wong, Z. Phys. {\bf C53} (1992) 465; Z. Phys. {\bf C58}
 		(1993) 159.
\item{[7]}	J.I.~Kapusta, P.~Lichard and D.~Seibert, Phys. Rev. {\bf D44}
		(1991) 2774.
\item{[8]}	R.~Baier, H.~Nakkagawa, A.~Ni\'egawa and K.~Redlich, Z. Phys.
		{\bf C53} (1992) 433.
\item{[9]}	I.~Matsubara, Prog. Theor. Phys. {\bf 14} (1993) 351.
\item{[10]} 	A.~Fetter and J.~Walecka, Quantum theory of many particle
		systems, Mc Graw Hill, 197.
\item{[11]}	A.~Niemi and G.W.~Semenoff, Ann. Phys. {\bf 152} (1984) 105; \
		Nucl. Phys. {\bf B230} (1984) 181.
\item{[12]}	R.L.~Kobes and G.W.~Semenoff, Nucl. Phys. {\bf B260} (1985)
		714; \ Nucl. Phys. {\bf B272} (1986) 329.
\item{[13]}	H.~Umezawa, H.~Matsumoto and M.~Tachiki, Thermo-field dynamics
		and condensed states, North-Holland, Amsterdam 1982.
\item{[14]}	R.~Kobes, Phys. Rev. {\bf D42} (1990) 562;\
		Phys. Rev. {\bf D43} (1991) 1269.
\item{[15]}	R.~Kobes, Phys. Rev. Lett. {\bf 67} (1991) 1384.
\item{[16]}	T.S.~Evans, Phys. Lett. {\bf B252} (1990) 108;\
		Nucl. Phys. {\bf B374} (1992) 340.
\item{[17]}	P.~Aurenche, T.~Becherrawy, Nucl. Phys. {\bf B379} (1992) 259.
\item{[18]}	C.M.A.~van~Eijck and Ch.G.~van~Weert, Phys. Lett. {\bf B278}
		(1992) 305.
\item{[19]}	P.~Aurenche, E.~Petitgirard and T.~del~Rio~Gaztelurrutia, Phys.
		Lett. {\bf B297} (1992) 337.
\item{[20]}	F.~Gu\'erin, Institut non Lin\'eaire de Nice preprints,
		INLN-1991/11; INLN-1993-14, May 1993.
\item{[21]}	P.A.~Henning and H.~Umezawa, preprint GSI-92-61, September
		1992; Phys. Lett. {\bf B303} (1993) 209.
\item{[22]}	H.A. Weldon, Phys. Rev. {\bf D26} (1982) 2789.
\item{[23]}	V.V.~Klimov, Sov. Journ. Nucl. Phys. {\bf 33} (1981) 934;\
		JETP {\bf 55} (1982) 199; \hfil
\item{ }	H.A.~Weldon, Phys. Rev. {\bf D26} (1982) 199; \ Phys. Rev.
		{\bf D26} (1982) 1394.
\item{[24]}	E.S.~Fradkin, Proc. (Trudy) of  the P.N.~Lebedev, Physics
		Institute, vol. {\bf 29}, D.V.~Skobel'tsyn ed., Consultants
		bureau, New York, 1967 (Quantum Field Theory and Hydrodynamics).
\item{[25]}	T.~Altherr and P.V.~Ruuskanen, Nucl. Phys. {\bf B380} (1992)
		377.
\item{[26]}	R.~Baier, S.~Peign\'e and D.~Schiff,  preprint
		LPTHE-Orsay 93/46.
\item{[27]}	C.~Gale and J.I.~Kapusta, Nucl. Phys. {\bf B357} (1991) 65.
\item{[28]}	V.V. Lebedev and A.V.~Smilga, Physica {\bf A181} (1992)
		187;
\item{ }	T.~Altherr, E.~Petitgirard and T.~del~Rio~Gaztelurrutia, Phys.
		Rev. {\bf D47} (1993) 703;
\item{ } 	R.~Baier, H.~Nakkagawa and A.~Ni\'egawa, Can. J. Phys. {\bf 71}
		(1993) 205;
\item{ }	R.D.~Pisarski, Phys. Rev. {\bf D47} (1993) 5589.

\vfill\eject

{\bf Figure Captions}

\vskip .50cm

\noindent
\item{   Fig. 1} The general "real-time" contour in the complex time plane.
\item{   Fig. 2} The QED vertex in the R/A formalism :
		 a) with an incoming fermion and an incoming anti-fermion;
		 b) with an incoming fermion and an outgoing fermion.
\item{   Fig. 3} The generic two-point function.
\item{   Fig. 4} The generic three-point function.
\item{   Fig. 5} The three-point function as a sum of tree diagrams. The
		 {\bf /} on a
		 line means that the line is put on mass-shell and carries a
		 statistical factor: $\eps (l_{i0}) ({1 \over 2} + \eta_i
		 n^{[\eta_i]} (l_{i0})) \de (L_i^2)$.
\item{   Fig. 6} The fermion self-energy diagram.
\item{   Fig. 7} The photon-photon-quark-antiquark four-point function in QED.
\item{   Fig. 8} A diagram in the effective theory: the two-point function with
		 vertex corrections; a) one loop vertices; b) dressed vertices.
\item{   Fig. 9} Taking the discontinuity with respect to $p_0$ of a vertex up
		 to one loop. The cross represents the action of taking the
		 discontinuity (i.e. putting the corresponding line on shell).
\item{   Fig. 10}The soft fermion loop contribution to the rate of soft real
		 photon production in a QCD plasma.
\item{   Fig. 11}The same as Fig. 10 displaying the structure of the effective
		 vertices.

\vfill\eject

\end